\documentclass[aps,prl,10pt,twocolumn,a4paper,superscriptaddress, longbibliography]{revtex4-1}

\usepackage{amsmath}
\usepackage{amssymb}
\usepackage{mathtools}
\usepackage{physics}
\usepackage{textcomp}
\usepackage{upgreek}
\usepackage{units}
\usepackage[pdftex]{graphicx}
\usepackage{microtype}
\usepackage{enumitem}
\usepackage{xcolor}
\usepackage{gnuplottex}    
\usepackage{multirow}   
\usepackage{hyperref}
\usepackage{placeins}
\usepackage{siunitx}
\usepackage{csquotes}

\usepackage{txfonts}

\renewcommand{\vec}[1]{\ensuremath{\boldsymbol{#1}}}

\newcommand{\nk}{{n \vec{k}}}

\newcommand{\mub}{\mu_\mathrm{B}}

\usepackage{pifont}
\begin{document}

\title{
Spin Seebeck and Spin Nernst Effects of Magnons in Noncollinear Antiferromagnetic Insulators
}

\author{Alexander Mook}
\affiliation{Institut f\"ur Physik, Martin-Luther-Universit\"at Halle-Wittenberg, D-06099 Halle (Saale), Germany}

\author{Robin R. Neumann}
\affiliation{Institut f\"ur Physik, Martin-Luther-Universit\"at Halle-Wittenberg, D-06099 Halle (Saale), Germany}

\author{J\"urgen Henk}
\affiliation{Institut f\"ur Physik, Martin-Luther-Universit\"at Halle-Wittenberg, D-06099 Halle (Saale), Germany}

\author{Ingrid Mertig}
\affiliation{Institut f\"ur Physik, Martin-Luther-Universit\"at Halle-Wittenberg, D-06099 Halle (Saale), Germany}
\affiliation{Max-Planck-Institut f\"ur Mikrostrukturphysik, D-06120 Halle (Saale), Germany}

\begin{abstract}
Our joint theoretical and computer experimental study of heat-to-spin conversion reveals that noncollinear antiferromagnetic insulators are promising materials for generating magnon spin currents upon application of a temperature gradient: they exhibit spin Seebeck and spin Nernst effects. Using Kubo theory and spin dynamics simulations, we explicitly evaluate these effects in a single kagome sheet of potassium iron jarosite, KFe$_3$(OH)$_6$(SO$_4$)$_2$, and predict a spin Seebeck conversion factor of $\unit[0.2]{\muup V/K}$ at a temperature of $\unit[20]{K}$.
\end{abstract}

\date{\today}

\maketitle

\paragraph*{Introduction.}
Interconversion phenomena between physical quantities like sound, charge, spin, or heat \cite{Otani2017} are cornerstones in the solid-state research for next-generation alternatives to today's CMOS technology. Two particularly active fields are those of spinelectronics (charge to spin and \textit{vice versa}) \cite{Zutic2004} and spincaloritronics (heat to spin and \textit{vice versa}) \cite{bauer2012spin}. While the former relies on electrons, the latter may disregard electrons as fundamental carriers in favor of collective magnetic excitations, i.\,e., magnons, thereby circumventing Joule heating.

A prominent magnonic heat-to-spin conversion phenomenon, which promises temperature control as well as waste-heat recovery, is the spin Seebeck effect (SSE) \cite{Uchida2010}, comprising a spin current in a magnetic insulator as response to an applied temperature gradient.
Magnons that ``flow down'' the gradient carry spin from the hot to the cold side of the sample. Accumulated at these ends, the spin diffuses into an adjacent heavy metal layer and gets converted into a transverse charge current by the inverse spin Hall effect \cite{Saitoh2006}.

While the SSE is natural to ferromagnets, it does not show up in uniaxial collinear antiferromagnets, because of their spin-degenerate magnon bands. Only an external magnetic field, which causes a Zeeman splitting of the magnon bands, introduces nonzero spin Seebeck signals \cite{Ohnuma2013, Cheng2014, Brataas2015, Seki2015, Wu2016, Rezende2016AFM}. Thus, the \textit{status quo} is that the heat-to-spin conversion by means of the SSE is possible in either ferromagnets (e.\,g., LaY$_2$Fe$_5$O$_{12}$ \cite{Uchida2010}) or uniaxial collinear antiferromagnets or paramagnets (e.\,g., MnF$_2$ \cite{Wu2016} and GGG \cite{Wu2015}, respectively) in magnetic fields, or biaxial collinear antiferromagnets (e.\,g., NiO \cite{holanda2017spin}) with nondegenerate magnon bands in zero field.

An alternative to the SSE is offered by the spin Nernst effect (SNE), which describes a transverse spin current as a response to a temperature gradient in magnetic insulators. It is found both in ferromagnets \cite{Kovalev2016, Han2016ax, wang2018anomalous, Mook2018}, collinear antiferromagnets \cite{Cheng2016, Zyuzin2016, Shiomi2017, Mook2018}, and paramagnets \cite{Zhang2018}. However, its proportionality to the strength of spin-orbit coupling (SOC) renders the heat-to-spin conversion rather inefficient. Therefore, it is about time to consider spin transport in a different material class, namely in noncollinear antiferromagnetic insulators (NAIs).

Herein, we show that NAIs are, in principle, materials for the generation of bulk magnon spin currents in \emph{zero magnetic field} and \emph{without SOC}. Taking a single kagome sheet of the NAI potassium iron jarosite KFe$_3$(OH)$_6$(SO$_4$)$_2$ as an example, we identify spin Seebeck and planar spin Nernst signals due to in-plane polarized bulk spin currents both within Kubo transport theory as well as atomistic spin dynamics simulations. Using superordinate symmetry arguments, these SSEs and planar SNEs are established as the magnonic version of spin-polarized electron currents in noncollinear antiferromagnetic metals \cite{Zelezny2017, Kimata2019}.

\begin{figure}
    \centering
    \includegraphics[width=1\columnwidth]{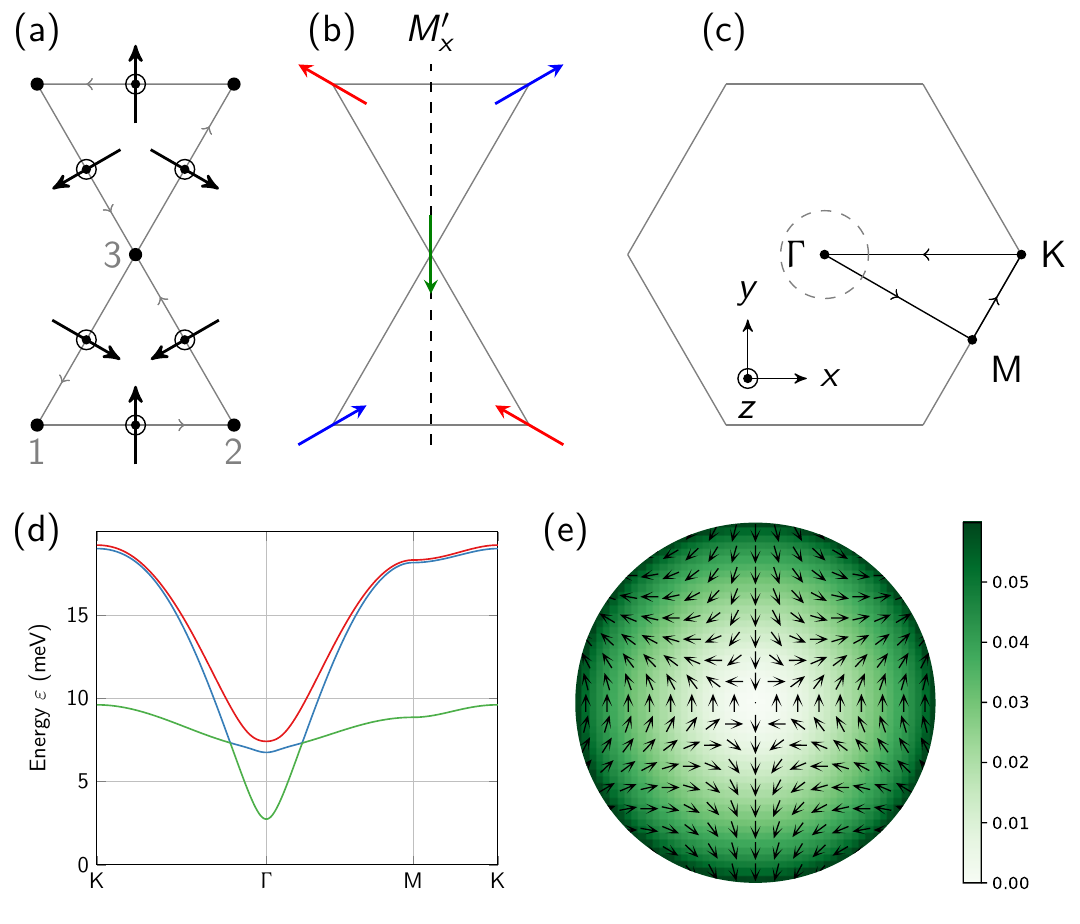}
    \caption{Single kagome layer of KFe$_3$(OH)$_6$(SO$_4$)$_2$. 
    (a) Structural lattice with three atoms per unit cell. DMI vectors are indicated as arrows for counterclockwise circulation. (b) In-plane components of the magnetic PVC order with $M'_x$ time-reversal mirror plane. (c) First Brioullin zone of the kagome lattice with indicated high symmetry points. (d) Magnon dispersion relation (at $B_z = \unit[-17]{T}$) along the path marked in (c). (e) Wave vector resolved spin expectation values of the lowest band in the vicinity of the $\Gamma$ point [see dashed circle in (c)]; color indicates the absolute value (in units of $\hbar$) and arrows the direction.}
    \label{fig:jarosite}
\end{figure}

\paragraph*{Heat-To-Spin Conversion in Potassium Iron Jarosite.}
We consider a single kagome sheet of potassium iron jarosite, KFe$_3$(OH)$_6$(SO$_4$)$_2$, which is an electrically insulating mineral built from Fe kagome planes [cf.~Fig.~\ref{fig:jarosite}(a)] stacked along the $c$ direction in ABC sequence \cite{Townsend1986}. The almost classical $S=5/2$ spins order below $\unit[65]{K}$ in the positive vector chiral (PVC) phase \cite{Inami1998, Inami2000} depicted in Fig.~\ref{fig:jarosite}(b). This phase is characterized by a positive $z$ component of the vector spin chirality
$
    \vec{\kappa} =  \vec{S}_1 \times \vec{S}_2 + \vec{S}_2 \times \vec{S}_3 + \vec{S}_3 \times \vec{S}_1  
$, 
where $\vec{S}_i$ ($i=1,2,3$) are the three spins in the magnetic unit cell as indicated in Figs.~\ref{fig:jarosite}(a) and (b). 

The two-dimensional (2D) spin Hamiltonian \cite{Elhajal2002,Yildirim2006}
\begin{align}
	\mathcal{H} = \frac{1}{2\hbar^2}
	\sum_{ ij } \left( J_{ij} \vec{S}_i \cdot \vec{S}_j + \vec{D}_{ij} \cdot \vec{S}_i \times \vec{S}_j \right) + \frac{g \mub}{\hbar} B_z \sum_i S_i^z \label{eq:spin-Ham}
\end{align}
includes antiferromagnetic exchange $J_{ij}$ between nearest ($\unit[3.18]{meV}$ \cite{Matan2006,Yildirim2006}) and next-nearest neighbors ($\unit[0.11]{meV}$ \cite{Matan2006,Yildirim2006}), which---due to geometric frustration---favors any classical $120^\circ$ ground state. The Dzyaloshinskii-Moriya interaction (DMI) \cite{Dzyaloshinsky58, Moriya60} between nearest neighbors is described by the vector $\vec{D}_{ij} = \pm (D_\parallel \hat{\vec{n}}_{ij} + D_z \hat{\vec{z}})$ [positive (negative) sign for counterclockwise (clockwise) circulation], possessing out-of-plane ($D_z = -0.062 J$ \cite{Matan2006,Yildirim2006}) as well as in-plane components ($D_\parallel = 0.062 J$ \cite{Matan2006,Yildirim2006}). The latter arise because the kagome planes are no mirror planes \cite{Elhajal2002}. $\vec{D}_{ij}$ is orthogonal to the $ij$ bond and $\hat{\vec{n}}_{ij}=\hat{\vec{n}}_{ji}$ is an in-plane unit vector as shown in Fig.~\ref{fig:jarosite}(a). A sign convention opposite to that of Ref.~\onlinecite{Elhajal2002} is used: $D_z < 0$ stabilizes the PVC phase \cite{Elhajal2002} and $D_\parallel$ causes a tiny out-of-plane canting \cite{Elhajal2002} (canting angle $1.98^\circ$ \cite{Matan2006,Yildirim2006}). Finally, we consider a magnetic field $\vec{B}$ along $z$ direction (with g-factor $g=2.13$ \cite{Grohol2005} and Bohr's magneton $\mub$). 

At zero magnetic field the spin textures of adjacent kagome layers in bulk KFe$_3$(OH)$_6$(SO$_4$)$_2$ are exactly opposite \cite{Inami1998, Inami2000} due to weak interlayer coupling ($\approx - \unit[0.03]{meV}$ \cite{Matan2011}) that is neglected here. Upon application of a sufficiently large magnetic field of $B_z \approx - \unit[17]{T}$ \cite{Matan2011, Fujita2012} the spin orientations in every second layer flip and the kagome sheets exhibit identical magnetic orders \cite{Grohol2005, Matan2011, Fujita2012}. As we show in the following, identical textures are important to ensure a finite spin current generation upon application of a temperature gradient. Thus, the results obtained for the 2D model at hand apply to the actual bulk material for $B_z \lesssim -\unit[17]{T}$.

Taking this magnetic field into account, we determine the resulting canting angle numerically ($\approx 2.85^\circ$) and carry out linear spin-wave calculations (cf.~Sec.~I A of the Supplemental Material (SM) \cite{Supplement}). We obtain the magnon energies $\varepsilon_\nk$ (with $n=1,2,3$) shown in Fig.~\ref{fig:jarosite}(d) along high symmetry lines depicted in Fig.~\ref{fig:jarosite}(c). Following Ref.~\onlinecite{Okuma2017}, we calculate the spin expectation values of magnons in the lowest band close to the Brillouin zone center [Fig.~\ref{fig:jarosite}(e)]. We find the double winding of the magnon spin direction about the $\Gamma$ point known from Ref.~\onlinecite{Okuma2017}. This spin-momentum locking suggests the possibility of net spin currents in nonequilibrium. 

We are interested in the magnetothermal transport tensor $\varUpsilon_{\mu \nu}^\gamma$, that mediates between the temperature gradient $\nabla_\nu T$ in $\nu$ direction and the nonequilibrium spin current density $\langle j_\mu^\gamma \rangle$ of $\gamma$ spin in $\mu$ direction: $\langle j_\mu^\gamma \rangle = \varUpsilon_{\mu \nu}^\gamma (- \nabla_\nu T)$. This tensor comprises the SSE (diagonal elements, $\mu=\nu)$, the SNE (off-diagonal elements, $\mu \ne \nu$; antisymmetric part of $\varUpsilon^\gamma$), and the planar SNE (symmetric part of $\varUpsilon^\gamma$). Applying Kubo's theory \cite{Kubo1957,Mahan2000,Matsumoto2014,Zyuzin2016} and considering only the intraband contributions proportional to a phenomenological magnon relaxation time $\tau$, we find (cf.~Sec.~I B of SM \cite{Supplement} for the derivation and a discussion of approximations)
\begin{align}
    \varUpsilon_{\mu\nu}^\gamma = \frac{\tau}{2VT} \sum_{\vec{k}} \sum_{n=1}^{2N} \mathrm{Re} \left[ \left( J_{\vec{k},\mu}^\gamma \right)_{nn} \right] \left( Q_{\vec{k},\nu} \right)_{nn}  \left( - \frac{\partial \rho}{\partial \varepsilon} \right).
	\label{eq:MagnetothermalTransportTensorKubo}
\end{align}
$(J_{\vec{k},\mu}^\gamma)_{nn}$ and $(Q_{\vec{k},\mu})_{nn}$ denote diagonal matrix elements of the spin and heat current operators, respectively. $\rho=[\exp(\beta (G \mathcal{E}_{\vec{k}})_{nn})-1]^{-1}$ is the Bose-Einstein distribution function, $\beta = (k_\mathrm{B} T)^{-1}$ with $k_\mathrm{B}$ denoting Boltzmann's constant, and $V$ is the sample's volume. $G$ is the bosonic metric and $\mathcal{E}_{\vec{k}}$ the paraunitarily diagonalized Hamiltonian, containing the magnon energies (cf.~Sec.~I A of SM \cite{Supplement}).

Eq.~\eqref{eq:MagnetothermalTransportTensorKubo} describes the time-odd part of the full magnetothermal transport tensor (cf.~Sec.~I B of SM \cite{Supplement}). To see so, recall that $J_{\vec{k},\mu}^\gamma$ is even but $Q_{\vec{k},\mu}$ odd under time reversal. Thus, reversal of the magnetic texture reverses the sign of $\varUpsilon_{\mu\nu}^\gamma$, which is well-known for the SSE in ferromagnets \cite{Uchida2010}. This also explains the absence of the SSE in those antiferromagnets for which time reversal can be ``repaired'' by a sublattice swap: such antiferromagnets are effectively time-even and as such incompatible with a time-odd transport response. In zero field, bulk KFe$_3$(OH)$_6$(SO$_4$)$_2$ exhibits such a symmetry due to the opposite spin textures of adjacent kagome sheets and we expect zero $\varUpsilon$. This is why we consider the spin-flopped phase with $B_z < -\unit[17]{T}$ modeled by a single layer.

\begin{figure}
    \centering
    \includegraphics[width=\columnwidth]{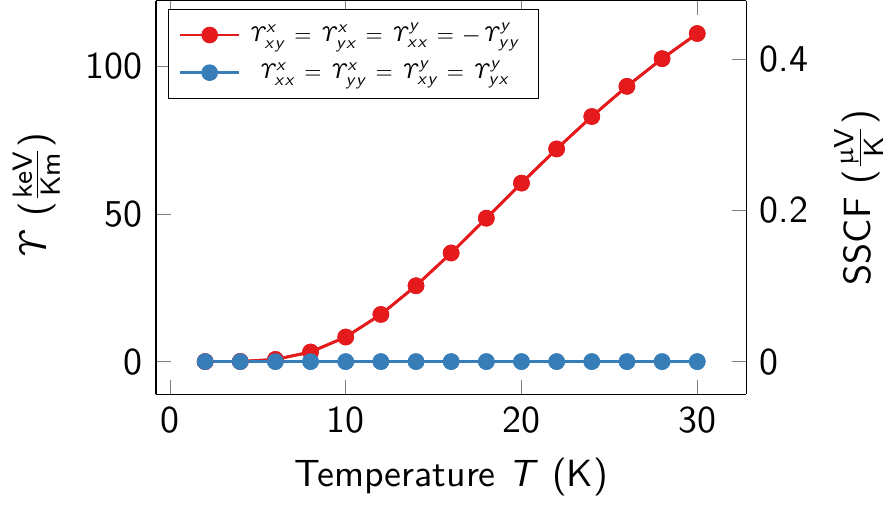}
    \caption{Temperature dependence of $\varUpsilon^\gamma_{\mu\nu}$ ($\mu,\nu,\gamma = x,y$) in KFe$_3$(OH)$_6$(SO$_4$)$_2$ for $B_z = \unit[-17]{T}$. Left ordinate: natural units of $\varUpsilon$ in three dimensions. Right ordinate: corresponding spin Seebeck conversion factor (SSCF; inverse spin Hall voltage divided by temperature gradient) in an adjacent platinum layer.}
    \label{fig:cond_jarosite_sni_hf}
\end{figure}

Applying Eq.~\eqref{eq:MagnetothermalTransportTensorKubo} to the 2D model of KFe$_3$(OH)$_6$(SO$_4$)$_2$, we calculate the elements $\varUpsilon^\gamma_{\mu\nu}$ for $\mu,\nu,\gamma = x,y$ (in-plane transport of in-plane polarized spins); results are shown in Fig.~\ref{fig:cond_jarosite_sni_hf} \footnote{The results in Fig.~\ref{fig:cond_jarosite_sni_hf} were obtained for a realistic effective magnon relaxation lifetime $\tau = \unit[0.1]{ns}$. We divided the result of the two-dimensional integral in Eq.~\eqref{eq:MagnetothermalTransportTensorKubo} by the kagome interlayer distance of $\unit[0.57]{nm}$ \cite{Townsend1986, Grohol2003} to obtain three-dimensional units.}. There are several nonzero elements, which are identical in modulus (red line); the transport tensor assumes the form
\begin{align}
	\varUpsilon^x = \begin{pmatrix}
		0 & \varUpsilon^x_{xy}\\
		\varUpsilon^x_{xy} & 0
	\end{pmatrix},
	\quad
	\varUpsilon^y = \begin{pmatrix}
		\varUpsilon^x_{xy} & 0 \\
		0 & -\varUpsilon^x_{xy}
	\end{pmatrix}. \label{eq:PVC}
\end{align}
Consequently, when applying $\vec{\nabla}T$ in $x$ or $y$ direction, there is a longitudinal magnon particle current density, consisting of $y$ spin-polarized magnons (diagonal elements of $\varUpsilon^y$): there is a SSE. Moreover, there is a transverse $x$-polarized spin current (off-diagonal elements of $\varUpsilon^x$), i.\,e., a planar SNE\@.
At $\unit[20]{K}$ we find a spin Seebeck coefficient of about $\unit[60]{keV/(Km)}$ (left ordinate in Fig.~\ref{fig:cond_jarosite_sni_hf}). Assuming an inverse-spin-Hall spin-to-charge current conversion factor of $\unit[1.3 \times 10^{-4}]{Vs/(Am)}$ \cite{Basso2016} in an adjacent platinum layer, this corresponds to a spin Seebeck conversion factor (SSCF) of $ \unit[0.2]{\muup V/K}$ (right ordinate in Fig.~\ref{fig:cond_jarosite_sni_hf}). This is to be compared with values of ferrimagnetic YIG ($\lesssim \unit[5]{\muup V/K}$ \cite{Kikkawa2015}) or of collinear antiferromagnets in magnetic fields like Cr$_2$O$_3$ ($\unit[0.015]{\muup V/K}$ at $\unit[14]{T}$ and $\unit[35]{K}$ \cite{Seki2015}), MnF$_2$ ($\unit[41.2]{\muup V/K}$ at $\unit[14]{T}$ and $\unit[15]{K}$ \cite{Wu2016}), or $\alpha$-Cu$_2$V$_2$O$_7$ ($\unit[0.1]{\muup V/K}$ at $\unit[5]{T}$ and $\unit[2]{K}$ \cite{Shiomi2017SSE}). Hence, the SSE in KFe$_3$(OH)$_6$(SO$_4$)$_2$ is well within experimental range; the same arguments apply to the planar SNE\@.

In contrast to the aforementioned collinear antiferromagnets that exhibit the SSE only in magnetic fields, we also obtain a SSE in \emph{zero magnetic field} (cf.~Sec.~II of SM \cite{Supplement}), which would be experimentally accessible in a single kagome sheet.

\paragraph*{Computer experiment.} 
Since spin is not conserved, spin currents are detected indirectly by measurement of the observable spin accumulation they bring about in samples with finite dimensions; standard means include the inverse spin Hall effect in an adjacent normal metal layer \cite{Saitoh2006}, spin torque in an adjacent ferromagnet \cite{Locatelli2013}, or magnetooptical Kerr microscopy \cite{Kato2004}. We now demonstrate that the SSE and planar SNE in NAIs cause finite nonequilibrium spin accumulations.

\begin{figure}
\centering
\includegraphics[width=\columnwidth]{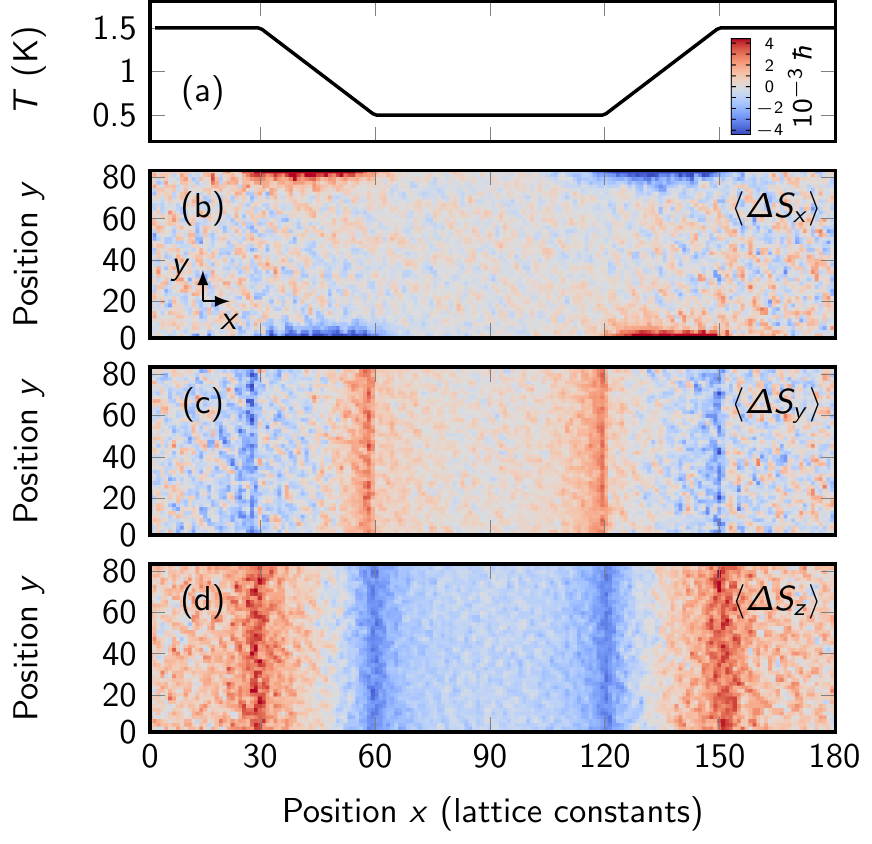}
\caption{Nonequilibrium spin accumulation in a single kagome sheet of KFe$_3$(OH)$_6$(SO$_4$)$_2$ as obtained by spin dynamics simulations. (a) Temperature profile with two opposite gradients. (b)--(d) Position-resolved $x$, $y$, and $z$ components of the nonequilibrium spin accumulation $\langle \Delta \vec{S} \rangle$. While $x$ spin accumulates at the sample's edges [SNE geometry, (b)], $y$ and $z$ spin accumulates in longitudinal direction [SSE geometry, (c) and (d)], i.\,e., at the ends of the temperature gradient. Opposite temperature gradients cause opposite accumulations. Red/white/blue color indicates positive/zero/negative accumulation.}
\label{fig:Simulation}
\end{figure}

The actual experimental situation, in which a temperature gradient is applied to the magnet, is simulated by relying on classical atomistic spin dynamics simulations based on the stochastic Landau-Lifshitz-Gilbert equation \cite{evans2014atomistic}. We set up a rectangular stripe of a single KFe$_3$(OH)$_6$(SO$_4$)$_2$ kagome sheet (built from about $50\,000$ spins) with finite width in $y$ direction and periodic boundary conditions along the longer $x$ direction. The orientation of the kagome triangles and the spin ordering is as indicated in Fig.~\ref{fig:jarosite}(b). Each spin is coupled to its own heat bath at a spatially varying temperature as shown in Fig.~\ref{fig:Simulation}(a); the temperature profile exhibits two opposite gradients in $x$ direction. Then, inspired by Ref.~\onlinecite{Ritzmann2014}, we measure a position-resolved steady-state nonequilibrium spin accumulation $\langle \Delta \vec{S}_i \rangle$ (cf.~Sec.~III of SM \cite{Supplement} for technical details).

The position-resolved $x$, $y$, and $z$ components of $\langle \Delta \vec{S}_i \rangle$ are shown in panels (b), (c) and (d) of Fig.~\ref{fig:Simulation}, respectively. In panel (b), an accumulation of $x$ spin is observed at the edges of the sample in those regions with a finite $\nabla_x T$ (cf.~red and blue horizontal stripes), indicating a transverse spin current as expected from the off-diagonal elements of $\varUpsilon^x$ [cf.~Eq.~\eqref{eq:PVC}]. In contrast, there is zero $x$ spin accumulation at the ends of the gradients, which is in agreement with zero diagonal elements of $\varUpsilon^x$. Overall, Fig.~\ref{fig:Simulation}(b) proves the existence of the planar SNE with $x$-polarized transverse spin currents.

In Fig.~\ref{fig:Simulation}(c), a finite $y$ spin accumulation is observed at the ends of the thermal gradients (cf.~red and blue vertical stripes), which is in accord with the nonzero diagonal elements of $\varUpsilon^y$ [cf.~Eq.~\eqref{eq:PVC}]. However, no $y$ spin accumulates at the edges of the sample (zero off-diagonal elements of $\varUpsilon^y$). These results demonstrate the existence of the SSE with $y$-polarized longitudinal spin currents. 

We also observe a longitudinal accumulation of $z$ spin in Fig.~\ref{fig:Simulation}(d). It is caused by the small out-of-plane canting induced by the in-plane DMI, due to which all magnons acquire a spin component in $z$ direction (cf.~Sec.~IV A of SM \cite{Supplement}). This effect is just the usual SSE associated with ferromagnetism.

In Sec.~III D of the SM \cite{Supplement} we show that reversal of the magnetic texture leads to a sign reversal of spin accumulations, unambiguously relating the accumulations with the time-odd part of $\varUpsilon$, which is in accordance with theory. This finding corroborates further that spin current responses of kagome lattices with opposite textures cancel out.

\paragraph*{Origin of the SSE and SNE\@.}
The qualitative results we have obtained here, namely the existence of both a SSE and SNE, are not limited to the material under consideration. They equally apply to any kagome antiferromagnet in the PVC phase as is evident from superordinate symmetry considerations.

The slightly out-of-plane tilted PVC phase has a three-fold rotational axis pointing out of the plane ($C_{3z}$) and a $M'_x$ time-reversal mirror plane whose normal points along $x$ [Fig.~\ref{fig:jarosite}(b)]. Applying Neumann's principle \cite{neumann1885vorlesungen}, i.\,e., requiring the magnetothermal transport tensor to be invariant under the magnetic crystal symmetries, the shape of $\varUpsilon$ in Eq.~\eqref{eq:PVC} can be derived rigorously (cf.~Sec.~IV A of SM \cite{Supplement}). In essence, the SSE and planar SNE are allowed to exist, because the symmetry of the noncollinear PVC texture is too low to forbid spin currents.

This finding is in accordance with the symmetry-restricted spin transport tensor shapes studied in Ref.~\onlinecite{Seemann2015} and the electronic spin-polarized currents in noncollinear antiferromagnetic metals \cite{Zelezny2017, Kimata2019} \footnote{We recall that we have restricted our focus to the time-odd part of $\varUpsilon$ given in Eq.~\eqref{eq:MagnetothermalTransportTensorKubo} (Sec.~I B of SM \cite{Supplement}); thus, the transport tensor expressions we list here do not appear in their full symmetry-allowed shape \cite{Seemann2015}.}.

In the Introduction we promised spin currents \emph{in zero field} and \emph{without SOC}. Armed with the above symmetry arguments, we construct a \emph{gedanken} magnet that keeps these promises. Starting from a KFe$_3$(OH)$_6$(SO$_4$)$_2$ sheet, the limit of zero SOC is obtained by setting the DMI zero. This results in a perfectly coplanar PVC phase (zero magnetization) stabilized by the second-nearest neighbor exchange \cite{Harris1992, Chubukov1992, Chernyshev2015largeS}. In addition to the $C_{3z}$ and $M'_x$ symmetries, there are now a $M_y$ and a $M'_z$ symmetry. The latter are still insufficient to forbid spin-polarized currents (Sec.~IV B of SM \cite{Supplement}) and the form of $\varUpsilon$ remains as in Eq.~\eqref{eq:PVC}. Thus, assuming a single kagome sheet with a single magnetic domain, we obtain both an SSE and SNE in zero field \footnote{In actual experiments, one might need a tiny in-plane magnetic field to energetically favor one particular domain.}.

We have numerically verified the SSE and planar SNE in the \emph{gedanken} magnet by calculating $\varUpsilon$ by Eq.~\eqref{eq:MagnetothermalTransportTensorKubo} (Sec.~IV B of SM \cite{Supplement}). There are three Goldstone modes (cf.~Fig.~S3 of SM \cite{Supplement}) \cite{Harris1992, Chubukov1992}, associated with a global rotation of the texture; all three magnon branches contribute to transport (cf.~Fig.~S4 of SM \cite{Supplement}).
Unfortunately, we are not aware of a material which strictly realizes this model, owing to the inevitibility of nearest-neighbor DMI on the kagome lattice. Since a single KFe$_3$(OH)$_6$(SO$_4$)$_2$ kagome sheet deviates only slightly from the \emph{gedanken} magnet, single-crystalline bulk KFe$_3$(OH)$_6$(SO$_4$)$_2$ \cite{Grohol2005} serves as a candidate material on which a proof-of-principle experiment can be performed. Other iron jarosites \cite{Grohol2003} are also candidates, e.\,g., silver iron jarosite AgFe$_3$(OH)$_6$(SO$_4$)$_2$, which orders below $\unit[59]{K}$ \cite{Matan2011} and takes $B_z < - \unit[10]{T}$ \cite{Matan2011} to ensure identical layer spin textures.

\paragraph*{Other antiferromagnetic textures.}
By extending the symmetry considerations to other noncollinear antiferromagnetic textures, it becomes evident that the SSE and planar SNE are inherent phenomena in NAIs.

In contrast to the PVC phase, the negative vector chiral (NVC) phase (cf. Fig.~S5 of SM \cite{Supplement}), which is stabilized for $D_z > 0$ \cite{Elhajal2002} (recall opposite sign convention) and characterized by $\kappa_z<0$, does not have a $C_{3}$ axis. Thus, symmetry considerations are restricted to an $M'_x$ time-reversal mirror plane, yielding (Sec.~IV C of SM \cite{Supplement})
\begin{align}
	\varUpsilon^x = \begin{pmatrix}
		0 & \varUpsilon^x_{xy} \\
		\varUpsilon^x_{yx} & 0
	\end{pmatrix},
	\quad
	\varUpsilon^y = \begin{pmatrix}
		\varUpsilon^y_{xx} & 0 \\
		0 & \varUpsilon^y_{yy}
	\end{pmatrix}. \label{eq:NVC}
\end{align}
Since in general $\varUpsilon^y_{xx} \ne \varUpsilon^y_{yy}$ and $\varUpsilon^x_{xy} \ne \varUpsilon^x_{yx}$, there is on top of the planar SNE of $x$-polarized spin (symmetric part of $\varUpsilon^x$), also a ``magnetic'' \footnote{The adjective ``magnetic'' is inspired by the magnetic spin Hall effect studied in Refs.~\onlinecite{Zelezny2017}, \onlinecite{Chen2018} and \onlinecite{Kimata2019}.} SNE (antisymmetric part of $\varUpsilon^x$). A $90^\circ$ rotated version of the NVC phase appears in cadmium kapellasite due to local anisotropies \cite{Okuma2017InverseChiral} (Sec.~IV D of SM \cite{Supplement}).

Concerning three-dimensional NAIs, we note that pyrochlores with the all-in--all-out texture as, e.\,g., Sm$_2$Ir$_2$O$_7$ \cite{Donnerer2016}, are expected to exhibit a planar SNE with the property that the force, the current, and the transported spin are mutually orthogonal to each other (Sec.~IV E of SM \cite{Supplement}).

\paragraph*{Conclusion.}
We showed that NAIs can exhibit bulk magnon spin currents, thus complementing the recent proposal of an interfacial SSE \cite{Flebus2018}. As results, NAIs offer the combined advantages of nonelectronic spin transport and antiferromagnetism. They may replace ferromagnets as spin-active components of next-generation spin(calori)tronic devices and introduce a paradigm of an ``antiferromagnetic insulator spincaloritronics'', as the magnonic pendant to the thriving field of ``antiferromagnetic spinelectronics'' \cite{Baltz2018, jungwirth2018multiple, gomonay2018antiferromagnetic, duine2018synthetic, vzelezny2018spin, nvemec2018antiferromagnetic, vsmejkal2018topological, jungfleisch2018perspectives}. 

Besides an experimental proof of principle of our quantitative predictions for KFe$_3$(OH)$_6$(SO$_4$)$_2$ our work calls for the development of a theory of magnon spin and heat diffusion \cite{Cornelissen2016} in NAIs, an investigation of the influence of noncollinearity-induced magnon-magnon interactions \cite{Chernyshev2015largeS} on spin transport, of the dynamics of noncollinear antiferromagnetic domain walls \cite{Lhotel2011} in temperature gradients, and a material search for NAIs with ordering temperatures above room temperature.

By virtue of Onsager's reciprocity relation \cite{Onsager1931}, the existence of SSEs and SNEs in noncollinear antiferromagnets, immediately implies that of the spin Peltier and spin Ettings\-hausen effects. A recent experimental proof \cite{Sola2018} of the reciprocity between the spin Seebeck and spin Peltier effect in a YIG/Pt bilayer could be conceptually carried over to noncollinear antiferromagnets.

\acknowledgements
This work is supported by SFB 762 of Deutsche Forschungsgemeinschaft (DFG).

\bibliography{short,newrefs}

\end{document}